\documentclass[twocolumn,twoside,slac_two]{revtex4}
\usepackage{graphicx}
\usepackage{fancyhdr}
\newcommand{\abs}[1]{\ensuremath{\left|#1\right|}}
\newcommand{\DAFNE}{DA\char8NE}
\newcommand{\order}[1]{\ensuremath{\cal O}(#1)}
\newcommand{\Fig}[1]{Fig.~\ref{#1}}
\newcommand{\Lcms}{\ensuremath{{\rm cm}^{-2}\,{\rm s}^{-1}}}
\newcommand{\Ref}[1]{Ref.~\onlinecite{#1}}
\newcommand{\Reps}{\ensuremath{{\rm Re}\ \varepsilon'/\varepsilon}}

\newcommand{\SN}[2]{\ensuremath{#1\times10^{#2}}}

\renewcommand{\Im}[1]{\ensuremath{\rm Im}\:#1}
\renewcommand{\Re}[1]{\ensuremath{\rm Re}\:#1}

\begin{document}

\title{New Directions in Kaon Physics}

\author{Matthew Moulson}
\affiliation{Laboratori Nazionali di Frascati dell'INFN, 
00044 Frascati (Roma), Italy\\
E-mail: \url{moulson@lnf.infn.it}}

\begin{abstract}
Recent measurements of kaon decays provide new information about 
CKM unitarity, lepton universality, 
and discrete symmetries. KLOE-2, the proposed extension of the 
kaon physics program at Frascati, will extend the world data set 
on kaon decays and conduct interference measurements with neutral kaons.
Meanwhile, the decays $K\to\pi\nu\bar{\nu}$ can be directly related to 
the CKM matrix elements with minimal theoretical uncertainty, and are 
the focus of a series of experiments.
Several events of $K^+\to\pi^+\nu\bar{\nu}$
have been observed to date; the goal of the NA62 experiment at CERN is 
to perform an \order{100}-event measurement in this channel. 
Initiatives in Japan---the E391a experiment at KEK, to become E14 at 
J-PARC---are focused on collecting a few $K_L\to\pi^0\nu\bar{\nu}$ events
in a first step, while working towards an \order{100}-event measurement.
Experiments capable of performing \order{1000}-event measurements in both
channels have been discussed.
\end{abstract}

\maketitle

\thispagestyle{fancy}

\section{Current Trends in Kaon Physics}
An abundance of new measurements of kaon decays are providing
precision tests of the flavor sector of the Standard Model (SM).
If the couplings of the $W$ to quarks and leptons are indeed specified
by a single gauge coupling, then the CKM matrix must be unitary 
for universality to be observed as the equivalence of the Fermi 
constant $G_F$ as measured in muon decay and in hadron decays. 
The current results from the FlaviaNet Kaon Working 
Group \cite{FlaviaNet+08:KSM} based on world data on $V_{us}$ from 
$K_{\ell3}$ decays and $V_{us}/V_{ud}$ from $K_{\ell2}$ decays (using
recent lattice QCD results \cite{UKRBC+07:f0,HPUK+07:fKpi} in either case), 
as well as on the world-average value of $V_{ud}$ from $0^+\to0^+$ nuclear 
$\beta$ decays \cite{TH07:Vud}, verify first-row CKM unitarity to within
$1 - \abs{V_{ud}}^2 - \abs{V_{us}}^2 = 0.0002(7)$, 
giving the second most precise determination of $G_F$ after that 
from muon decay.
From dimensional arguments,
this agreement can be said to constrain physics models mediated by
new particles ($Z'$ bosons, SUSY, technicolor) at mass scales of 
\order{1~{\rm TeV}} at the loop level, or \order{10~{\rm TeV}} at tree level
\cite{Mar07:Kaon}.

$K_{\ell2}$ decays are helicity suppressed. As a result,
hypothetical new-physics contributions are potentially observable 
in the $K_{\mu2}$ rate, such as
a tree-level contribution from the $H^+$ in certain two-Higgs-doublet models
(in analogy to the case of $B\to\tau\nu$; see \Ref{Hou93:Bt2}),
or contributions from right-handed quark currents \cite{B+07:RHq}.
However, evaluation of the rate in the SM is tricky; the uncertainty 
on $f_K$, the kaon decay constant, dominates $\Gamma_{\rm SM}(K_{\mu2})$.
FlaviaNet \cite{FlaviaNet+08:KSM} performs a fit to the experimental data on 
$\Gamma(K_{\mu2})/\Gamma(\pi_{\mu2})$, $V_{ud}$ from $0^+\to0^+$ decays, and
$V_{us}$ from $K_{\ell3}$ decays, as well as the recent lattice QCD values
for $f_K/f_\pi$ and $f_+(0)$, to obtain a value for $R^2_{K\mu2} \equiv 
\Gamma(K_{\mu2})/\Gamma_{\rm SM}(K_{\mu2})$ 
(the fit assumes first-row CKM unitarity as well).
The result of this analysis, $R_{K\mu2} = 1.004(7)$, can be used to 
exclude a region of the $\tan \beta$ vs.\ $m_{H^+}$ parameter space 
that is not excluded by $B\to\tau\nu$ data.

Because of the strong helicity suppression of the $K_{e2}$ decay,
its rate is particularly sensitive to new-physics
contributions. In minimal supersymmetric SM extensions (MSSM)
with $R$ parity, the $H^+$-mediated process with an effective 
$eH^+\nu_\tau$ coupling (dependent on the value of the mass insertion
$\Delta_{13}$) may add a lepton-flavor-violating contribution of 
as much as $\sim$1\% to the $K_{e2}$ rate, for $\tan \beta \sim 40$, 
$m_{H^+} \sim 500$~GeV, and $\Delta_{13} \sim \SN{5}{-4}$ 
\cite{MPP06:Ke2,MPP08:Ke2rev}.
In addition, observable deviations of $R_K$ from the SM value are expected in 
certain supersymmetric grand-unification models \cite{ELR08:Kl2GUT}. 
In the SM, the ratio $R_K = \Gamma(K_{e2})/\Gamma(K_{\mu2})$
has been calculated with precision: $R_K^{\rm SM} = \SN{2.477(1)}{-5}$
\cite{CR07:Kp2}.
The 2007 FlaviaNet experimental average (including preliminary results
from NA48 and KLOE) is $R_K = \SN{2.457(32)}{-5}$ \cite{FlaviaNet+08:KSM}.
There is much interest in reducing the experimental uncertainty on $R_K$. 

World data on $K_S$ and $K_L$ decay amplitudes and can be combined to 
to obtain refined values for the $CP$- and $CPT$-violation parameters 
$\Re{\varepsilon}$ and $\Im{\delta}$ using the Bell-Steinberger
unitarity relation. 
Such an analysis was performed in 2001 by the CPLEAR collaboration
\cite{CPLEAR+99:BSR}. Missing amplitude data for semileptonic 
$K_S$ decays were supplied by KLOE measurements \cite{KLOE+06:KSe3}, 
and in 2006, KLOE performed an improved unitarity analysis \cite{KLOE+06:BSR}.
At present, the uncertainties on the phases of the 
$K_L$-to-$K_S$ amplitude ratios 
$\eta_{+-}$ and $\eta_{00}$ limit the precision of the 
unitarity analysis.
Together with their recently announced final result for \Reps\  
\cite{Wor08:HQL}, KTeV quotes new measurements of the phase
differences $\phi_\varepsilon$ and $\Delta\phi$ (updating the results in
\Ref{KTeV+03:CPpar}), each with precision 
matching that of the current PDG fit, allowing further improvements.

Interference techniques can provide measurements of the amplitude ratios 
and other parameters of the $K_SK_L$ system, as well 
as a number of tests of quantum mechanics. At the Frascati
$\phi$ factory, \DAFNE, $K_SK_L$ pairs from $\phi$ decay are 
created in a pure $J^{PC} = 1^{--}$ quantum state. The relative decay time
distribution for $K_SK_L$ decays to final states $f_1f_2$ contains an
interference term that oscillates like $\cos(\Delta m\Delta t - \phi_{12})$,
where $\phi_{12}$ is the phase difference between the amplitude ratios
$\eta_{1,2}$ for $K_L$ and $K_S$ decays to each final state.
For $f_1=f_2=\pi^+\pi^-$, $\phi_{12}=0$, and no decays
are expected with $\Delta t = 0$, reflecting the antisymmetry of the initial
state contracted with the symmetry of the final state. KLOE does in fact
observe a deficit of decays at small $\Delta t$, confirming this prediction
of quantum mechanics. This observation can be quantified by multiplying 
the interference term in the fit function by $(1 - \zeta)$, with the 
decoherence parameter $\zeta = 0$ in quantum mechanics. 
Working in the $K_SK_L$ and $K^0\bar{K^0}$ bases, 
KLOE obtains the preliminary results
$\zeta_{SL} = 0.009\pm0.022_{\rm stat}$ and
$\zeta_{00} = \SN{(0.030\pm0.012_{\rm stat})}{-6}$ \cite{DiD08:HQL}.
With increased statistics, KLOE could perform a wide variety of
interference measurements.     

\section{A Bridge to the Future: KLOE-2}
As observed in the preceding section, new results in 
kaon physics continue to be announced by experiments such as KLOE, KTeV, 
and NA48. An extension of the KLOE program has been proposed
\cite{KLOE2+06:LoI,KLOE2+07:prop} to pursue
new, higher-statistics measurements
of many (if not all) of the observables discussed above.

\begin{figure}[ht]
\centering
\includegraphics[width=40mm]{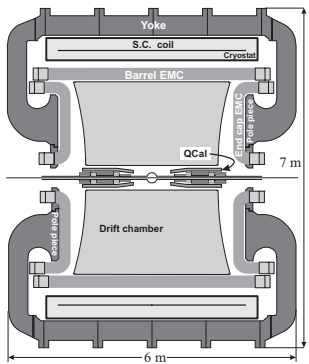}
\caption{Cross section of the KLOE detector.} 
\label{fig:kloe}
\end{figure}

The KLOE detector (\Fig{fig:kloe}) consists of a large ($r=2$~m) 
cylindrical drift chamber (DC),
surrounded by a lead/scintillating-fiber 
electromagnetic calorimeter (EmC), 
and features several optimizations for studies of the 
$K_SK_L$ system.
For a review of the KLOE detector and physics program, see \Ref{FM06:DAFrev}.
Most KLOE data were taken in 2001--2002 and in 2004--2005, for a total 
integrated luminosity of 2.5~fb$^{-1}$, or equivalently, of \SN{2.5}{9} 
$K_SK_L$ pairs produced.

The \DAFNE\ design luminosity of \SN{5}{32} \Lcms\ was never reached.
The peak \DAFNE\ luminosity in 2005 running
was about \SN{1.5}{32} \Lcms. 
The proposal to continue the KLOE physics program stems from the prospect 
for a \DAFNE\ upgrade to result in a substantial luminosity increase. 
To obtain high luminosity at \DAFNE\ requires a small value of the 
vertical beta function at the interaction point, $\beta^*_y$. 
Because $\beta^*_y$ can be made small only over a localized interval, this
in turn requires a short beam-overlap region. The solution adopted for
the \DAFNE\ upgrade \cite{RSZ07:CW} is the use of a beam-crossing 
angle that is large compared to the beam aspect ratio $\sigma_x/\sigma_z$.
This also reduces the effects of the beam-beam interaction and parasitic
collisions. An important element of the upgrade scheme is the use of 
a ``crabbed-waist'' sextupole correction, so that $\beta_y$ is always 
at a minimum at the collision point, regardless of the horizontal 
betatron coordinate, $x$.  This suppresses $x$-$y$ betatron and 
synchrobetatron resonances. The crabbed-waist interaction region has
generated much interest as a low-power-consumption solution for use at,
for example, a super $B$ factory or an upgraded LHC. At \DAFNE, 
the principle has been shown to work, and recently, a peak luminosity
of \SN{2.2}{32} \Lcms\ was obtained with beam currents somewhat lower
than those used in 2004--2005 running \cite{Rai08:EPAC}. \DAFNE\
performance continues to evolve rapidly.

\begin{figure}[ht]
\centering
\includegraphics[width=80mm]{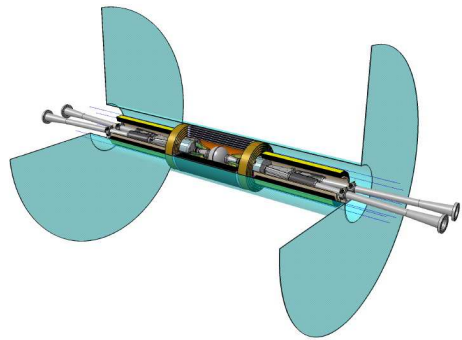}
\caption{Design of the KLOE-2 interaction region, showing the new inner
tracker and low-angle calorimeters.} 
\label{fig:kloe2}
\end{figure}

Two phases of KLOE-2 running are envisioned \cite{KLOE2+07:prop}. 
The first phase (``Step 0'') features 
minimal upgrades to the front-end electronics, DAQ, and online computing,
as well as necessary modifications to the interaction region to work
with new \DAFNE\ optics. Preparations will be complete in early 2009, 
and KLOE expects to collect about 5~fb$^{-1}$ before shutting down in 2010
to install an inner tracker and new forward calorimeters (\Fig{fig:kloe2}).
During the second phase of running (``Step 1''), KLOE expects to collect 
a data set of 20~fb$^{-1}$ or more.

The inner tracker will increase the acceptance for $K_S$ and 
$K^\pm$ decay products and improve the vertex resolution for
interferometry measurements. The goal is to improve the resolution on 
$\Delta t$, the time difference between $K_S$ and $K_L$ decays, from
$\sim$$0.9\,\tau_S$ (at present) to $0.25\,\tau_S$. At the same time, 
the material budget for the tracker must be held to an absolute minimum.
The solution chosen is a five-layer, cylindrical triple-GEM chamber
with a total material burden of less than $0.02\,X_0$. 
A single-layer prototype has been constructed and is operational
\cite{B+07:CGEM}. 

In addition, two new calorimeter systems will replace the current KLOE
quadrupole calorimeters. To improve the photon-veto efficiency at small 
angle (for measurements of decays such as $K_S\to\gamma\gamma$), LYSO
crystals will be placed in front of the new quadrupoles. LYSO
combines high light yield with excellent timing resolution.
To improve the hermeticity of the calorimeter system (e.g., to recognize
$K_L\to2\pi^0$ decays by photon veto), a lead/scintillating-tile 
calorimeter will be wrapped around the quadrupoles. The
$5\times5$~cm$^2$ tiles are individually read 
out by WLS fiber coupled to silicon photomultipliers, providing
high granularity \cite{KLOE2+08:cal}.

Some highlights of the Step-0 physics program at KLOE-2 include
measurements of the $K_{e2}$ branching ratio (BR) to 0.5\% and 
of the semileptonic $K_S$ BRs to 0.3--0.5\%, as well as 
a limit on ${\rm BR}(K_S\to3\pi^0)$ at 
the $10^{-8}$ level. The Step 1 upgrades will allow these measurements
to be further improved upon, and will also open the door to studies
of $CP$ and $CPT$ parameters in the $K_SK_L$ system and sensitive tests
of quantum mechanics using kaon interferometry \cite{Blo08:phipsi}.

\section{The Next Step: \boldmath{$K\to\pi\nu\bar{\nu}$}}
\begin{figure}[ht]
\centering
\includegraphics[height=60mm,angle=90]{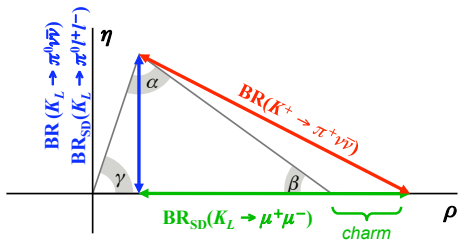}
\caption{Determination of the unitarity triangle with rare kaon decays.} 
\label{fig:ut}
\end{figure}

\begin{figure}[ht]
\centering
\includegraphics[width=80mm]{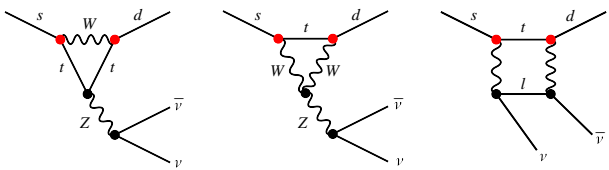}
\caption{Diagrams contributing to the process $K\to\pi\nu\bar{\nu}$.} 
\label{fig:fcnc}
\end{figure}

Four rare kaon decays provide information on the unitarity triangle,
as illustrated in \Fig{fig:ut}. 
These are flavor-changing neutral current
processes dominated by $Z$ penguin and box diagrams.
The $K\to\pi\nu\bar{\nu}$ decays proceed by the diagrams in \Fig{fig:fcnc}.
Their rates can be calculated with
minimal intrinsic uncertainty in the SM, because there are no contributions
from long-distance processes with intermediate photons, and because the 
hadronic matrix elements can be obtained from 
$K_{\ell3}$ rate and form factor measurements \cite{MS07:Kpnn}.
Because of the hierarchy of the CKM matrix elements and the hard GIM 
suppression for these processes, the rate for $K^+\to\pi^+\nu\bar{\nu}$
depends on the real and imaginary parts of $\lambda_t$, with 
some contribution from the real part of $\lambda_c$ 
($\lambda_q \equiv V_{qs}^*V_{qd}$).
A recent evaluation of the SM decay rate gives
${\rm BR} = \SN{(8.2\pm0.8)}{-11}$ \cite{FlaviaNet:webRare}. 
The principal contributions to the
error are the uncertainties on the current values of 
$\lambda_t$ and $m_c$. With no parametric uncertainties, the error
on the BR would be $\sim$7\%; the dominant non-parametric uncertainty
is from non-perturbative effects in the charm-loop and long-distance
contributions. (Recent work on the electroweak corrections to the
charm contribution may reduce the non-parametric error 
\cite{IMS05:Pcu,B+06:KpnnCharm,BG08:KpnnEWc}.)
The decay $K_L\to\pi^0\nu\bar{\nu}$ is $CP$ violating; its rate expression
depends on $\Im{\lambda_t}$. In this case, there are no contributions
to the uncertainty on the SM rate calculation from QCD corrections to 
the charm diagrams, and the prediction is particularly clean: 
${\rm BR} = \SN{(2.8\pm0.4)}{-11}$, with the uncertainties on 
$\lambda_t$ and $m_t$ as the principal error sources, and a 
non-parametric error of $\sim$3\% \cite{MS07:Kpnn}. As pointed out in
\Ref{Hai07:Kaon}, assuming future experiments measure both BRs to 10\%,
the determination of the unitary triangle from kaon decays would approach
the precision of that from $B$-meson decays. One hopes, however, to find 
signatures of new physics, which in many models are independent from 
those from $B$-meson observables. 
If there are indeed new sources of flavor-symmetry breaking at the TeV 
scale, such signatures may be dramatic. Although many proposed models 
have already been ruled out, an order-of-magnitude enhancement in
${\rm BR}(K_L\to\pi^0\nu\bar{\nu})$ is still possible in some MSSM 
\cite{B+05:MSSMrareK,I+06:MSSMrareK} and 
little-Higgs theories \cite{B+08:LHTrareK}. 
On the other hand, models incorporating
minimal flavor violation scenario feature small (20--30\%) deviations
in the expected rate. 
(For a recent review with references, see \Ref{Tar07:Kaon}.)
Either way, the measurement of the $K\to\pi\nu\bar{\nu}$ BRs with 10\%
precision would provide fundamental insight into the mechanism for
flavor-symmetry breaking, complementary to that to be obtained at the 
LHC. Experimentally, these measurements are extremely challenging.

\subsection{\boldmath{$K^+\to\pi^+\nu\bar{\nu}$}}
The main backgrounds to $K^+\to\pi^+\nu\bar{\nu}$ are $K^+\to\mu^+\nu$
(${\rm BR}=63\%$),
where the muon is identified as a pion, and  $K^+\to\pi^+\pi^0$
(${\rm BR}=21\%$), with two lost photons.
While these two-body decays can be kinematically recognized,
they must be rejected at the level of 10$^{-12}$.
In addition, there are backgrounds from
radiative decays (such as $K^+\to \mu^+\nu\gamma$ and $K^+\to\pi^+\pi^0\gamma$) 
and $K_{\ell3}$ and $K_{\ell4}$ decays, as well as from events in which 
beam particles ($\pi^+$, $K^+$) are scattered or interact in detector materials.
Excellent charged-particle identification (especially $\mu/\pi$ 
separation) and photon vetoes are critical. Experimental searches may
be conducted using either stopped kaons or decays in flight.

E787/949 at Brookhaven has actually measured this BR. The experiment 
makes use of a low-energy $K^+$ beam stopped in a scintillating-fiber
target. A DC surrounding the target measures the momentum ($B=1$~T)
of outgoing charged particles. $\pi^+$ identification is obtained in 
the range stack surrounding the DC, which is instrumented with custom
waveform digitizers capable of recording the decays at rest in the 
$\pi$-$\mu$-$e$ cascade. The experiment is surrounded by $4\pi$ of
photon vetoes. The main sequence of E787 running was from 1995--1998.
Two candidate events were recorded in a search window with the $\pi^+$
momentum above that for the $K^+\to\pi^+\pi^0$ decay. 
The experiment was upgraded in 2001 and began running as E949 in 2002. 
Originally scheduled for 60 weeks of running, the experiment was 
terminated after just 12 weeks due to funding cuts. 
E949 found an additional candidate
in the high momentum region, bringing the total number of candidates to three,
with an expected background of about 0.4 events.
The combined E787/E949 result for the high-momentum region is
${\rm BR} = \SN{(1.47^{+1.30}_{-0.89})}{-10}$ \cite{E949+08:Kpnn1rev}.
The collaboration has recently announced a new result for the search in the
lower momentum region. Three additional events were found, with an 
expected background of about 0.9 events.
Combining the data from both regions gives 
${\rm BR} = \SN{(1.73^{+1.15}_{-1.05})}{-10}$, in accordance with 
(though somewhat higher than) the SM prediction \cite{E949+08:Kpnn2}.

\begin{figure}[ht]
\centering
\includegraphics[width=72mm]{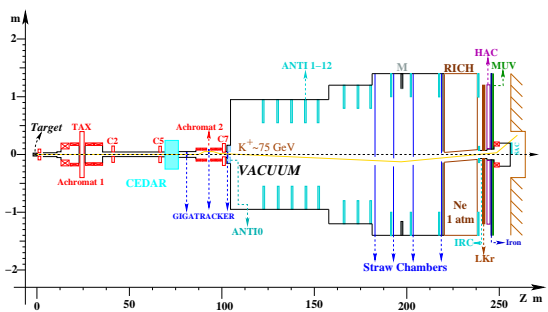}
\caption{Schematic diagram of the NA62 experiment. Note the different
vertical and horizontal scales.} 
\label{fig:na62}
\end{figure}

The next step is NA62 (also known as P-326), a proposed
decay-in-flight experiment at the CERN SPS \cite{P326+05:proposal}. 
The plan is to use elements of the NA48 apparatus 
to collect $\sim$100 $K^+\to\pi^+\nu\bar{\nu}$ decays
with a S/B ratio of 10:1 
in two years of operation \cite{P326+05:proposal}. 
The experimental layout is illustrated in \Fig{fig:na62} \cite{P326+07:update}.
The experiment will make use of a 75-GeV unseparated positive secondary beam.
The total beam rate is 800 MHz, providing $\sim$50 MHz of $K^+$'s.
After beam momentum selection and collimation, kaons are identified
in a differential Cerenkov counter (CEDAR) and the beam is 
tracked in the Gigatracker system, consisting of three thin 
silicon pixel detectors.
The decay volume begins 102~m downstream of the production 
target.   
About 5~MHz of kaon decays are observed in the first 60 m of the 120-m long 
vacuum decay region.
Large-angle photon vetoes are placed at 12 stations along 
the decay region and provide full coverage for decay photons with
$\mbox{8.5~mrad} < \theta < \mbox{50~mrad}$.
The last 35~m of the decay region hosts a dipole 
spectrometer ($\Delta p_\perp = 270$~MeV) with four straw-tracker 
stations operated in vacuum. 
A ring-imaging Cerenkov counter (RICH) downstream of the decay 
volume provides $\mu/\pi$ separation. Downstream of the RICH, 
the NA48 liquid-krypton
calorimeter (LKr) \cite{NA48+07:NIM}
is used to veto high-energy photons at small angle. 
Additional detectors further downstream extend the coverage
and efficiency of the muon detection and photon-veto systems.

$K^+\to\pi^+\pi^0$ decays must be rejected at the level of $10^{12}$.
Kinematic cuts on the $K^+$ and $\pi^+$ tracks provide a factor of $10^{4}$
and ensure 40~GeV of electromagnetic energy in the photon vetoes;
this energy must then be detected with an inefficiency of $10^{-8}$.
In $\sim$80\% of the events, both photons enter the forward veto;
in the remaining events, a hard photon enters the forward veto and 
a softer photon must be detected at large angle.
Studies of $K^+\to\pi^+\pi^0$ decays in NA48 data and tests conducted 
in 2006 with tagged photons from an $e^-$ beam confirm that the LKr
has an inefficiency of less than $10^{-5}$ for photons with 
$E_\gamma>10$~GeV, providing the needed rejection for forward photons
\cite{P326+07:update}.
The large-angle vetoes must have an inefficiency of $\lesssim$$10^{-4}$
for $E_\gamma$ as low as 200~MeV.
The plan is to construct the veto rings from lead-glass modules
from the OPAL electromagnetic calorimeter barrel \cite{OPAL+91:NIM}.
In a 2007 comparison of different technologies performed at Frascati,
the lead-glass modules were shown to have $10^{-4}$ inefficiency 
for 200~MeV electrons \cite{A+07:Veto}.

The particle-identification systems must 
provide $10^7$ $\mu$ rejection in $\pi$ selection. 
A standard downstream muon detector provides 
a factor of $10^5$; the additional rejection will come from a RICH
situated upstream of the LKr \cite{P326+07:update}.
In addition to providing $\mu$ suppression of better than $10^{-2}$ 
for $15<p<35$~GeV, the RICH must measure the track time with 
$\sigma_t=100$~ps and provide the charged-particle trigger.  
The RICH is a Ne-filled (1 atm) tube,
18 m long by 2.8~m in diameter,
with a through beam pipe and two inclined mirrors
to image the Cerenkov ring on 2000 compact PMTs 
packed in an 18-mm hexagonal lattice. 
A full-length prototype with 100 PMTs was tested in 2007 and provided 
$\sigma_t=75$~ps time resolution. A 400-PMT prototype will be tested in
fall 2008.

NA62 received preliminary approval for R\&D work by the CERN SPSC in 2006.
In 2007, NA62 collected more than $100\,000$ $K_{e2}$ decays with a subset
of the NA48/2 apparatus; a result for ${\rm BR}(K_{e2})$ with 0.5\%
precision is expected soon \cite{Spa08:HQL}. Detector prototypes were 
tested in 2007 running. More tests are scheduled for fall 2008, and 
the experiment hopes to obtain final approval by the end of the year. 
This would allow construction and installation in 2009--2011. Data
taking would begin in 2012. 

\subsection{\boldmath{$K_L\to\pi^0\nu\bar{\nu}$}}
The $K_L\to\pi^0\nu\bar{\nu}$ signature is the absence of
any detector activity other that from the two photons from the $\pi^0$.
All other $K_L$ final states have at least two extra photons or two charged
particles (except $K_L\to\gamma\gamma$, which is easy to reject kinematically).
The main $K_L$ background is from $K_L\to\pi^0\pi^0$ with two lost photons.
A hermetic and highly-efficient photon-veto system (including detectors
in the exit path of the neutral beam) is critical. Neutrons in the 
beam reacting with residual gas in the decay volume can also produce $\pi^0$s
(and $\eta$s), so the decay region must be kept under very high vacuum.
The reconstruction of the $\pi^0$ is the only sharp 
kinematic constraint available, and in beamline experiments, it is usually 
used to reconstruct the decay vertex position. Any successful experiment
must incorporate additional topological constraints by design, such as 
as a very small beam cross section (``pencil beam''), the measurement of the 
direction of incident photons, or the use of a microbunched beam to 
obtain time-of-flight constraints. (The latter two techniques were 
proposed for use in the KOPIO experiment \cite{KOPIO+99:prop}, which would
have run at Brookhaven but for which funding was cancelled in 2005.)

The first experiment dedicated to the search for $K_L\to\pi^0\nu\bar{\nu}$
is the E391a experiment at KEK \cite{K391+96:prop,K391+06:Run1,K391+08:Run2}.
\begin{figure}[ht]
\centering
\includegraphics[width=80mm]{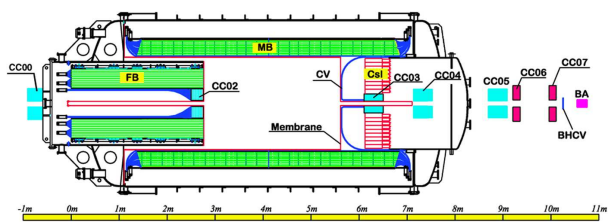}
\caption{Schematic diagram of the E391a experiment.}
\label{fig:e391a}
\end{figure}
The experiment makes use of a neutral secondary beam from the KEK PS,
collimated into a pencil beam.
The beam halo is suppressed to $10^{-4}$ at a radius of 4~cm. 
The $K_L$ momentum distribution at the detector peaks at 2~GeV.

The detector consists of two main regions: an upstream chamber, and
a larger vessel containing the fiducial decay region, as 
schematically illustrated in \Fig{fig:e391a}.
The upstream chamber is instrumented with the front barrel veto (FB),
which rejects beam halo and secondaries from $K_L$ decay upstream of
the fiducial volume. The main barrel veto (MB) covers the walls of the 
larger vessel, and encloses the fiducial decay volume, which is
evacuated to $10^{-7}$~mbar. Both the FB and MB vetoes are lead/scintillator 
sandwich counters. The main photon detector is an electromagnetic calorimeter 
at the downstream end of the decay volume, consisting of pure CsI crystals 
of $7\times7\times30$~cm$^3$. A plastic-scintillator hodoscope 50~cm
upstream of the calorimeter acts as a charged-particle veto (CV). 
A number of ring-shaped ``collar counters'' (CC) extend the coverage of 
the photon-veto system to small angles; the BA counter is a dual-readout
(scintillator/Cerenkov) sampling detector to veto photons leaving the 
experiment through the neutral-beam exit.

The experiment took data in three different runs in 2004--2005. 
Early results from the first run furnished the 90\% CL limit 
${\rm BR} \leq \SN{2.1}{-7}$ \cite{K391+06:Run1}, slightly 
improving on the previous limit from KTeV
(obtained in the channel in which $\pi^0\to e^+ e^-\gamma$)
\cite{KTeV+00:Kpnn}.
E391a has recently released the results of their
analysis of data from the second run \cite{K391+08:Run2}. 
Candidate $K_L\to\pi^0\nu\bar{\nu}$ events had two photons in the CsI and
no activity elsewhere in the detector. The longitudinal coordinate $z$ of the
$\pi^0$ vertex was obtained by assuming $M(\gamma\gamma) = m_{\pi^0}$,
with $M(\gamma\gamma)$ the two-photon invariant mass. A signal box was 
defined in the $p_\perp$ vs.\ $z$ plane; the estimated number of background
events in this box was $0.41\pm0.11$. No events were found, and an improved
90\% CL limit  ${\rm BR} \leq \SN{6.7}{-8}$ was obtained. 
The data sample from the third run is comparable in statistics to that 
from the second run, but the collaboration expects analysis improvements
to increase the signal acceptance.    

Much of the E391a detector will next be moved to the 
J-PARC facility nearing completion at Tokai.
Renamed E14, the next generation of the experiment aims to actually observe 
the $K_L\to\pi^0\nu\bar{\nu}$ decay \cite{JP14+06:prop}.
The neutral beam will be selected at 16$^\circ$ from a 30-GeV 
primary proton beam from the J-PARC main ring, and transported 
along a 20-m beamline in the Hadron Hall. The new beam will feature a 
core-to-halo ratio of $10^{-5}$ (down from $10^{-4}$ in E391a)
and a neutron to $K_L$ ratio of 7 (down from 40). The momentum 
spectrum at the experiment is slightly softer, which decreases
background levels from neutron interactions.
Among the upgrades to the detector, one of the most significant is the 
replacement of the E391a CsI crystals with those from the KTeV 
calorimeter \cite{KTeV+03:CPpar}.
The latter are both longer (50~cm) and smaller in cross section
($2.5\times2.5$ or  $5\times5$~cm$^2$). The higher granularity of the 
calorimeter will reduce $\gamma\gamma$ merging, while the increased depth
will reduce photon punch-through and longitudinal leakage. In particular,
$\pi^0$s from neutron interactions in the upstream detectors contribute 
background when their $z$ coordinates are misreconstructed because leakage
causes one of the photon energies to be mismeasured. This effect was 
measured in E391a by placing an aluminum plate in the beam to produce
$\pi^0$s; MC studies suggest that this background source will be 
substantially reduced in E14. 
Various other upgrades are planned as well, including a new 
lead/aerogel in-beam downstream photon veto (styled after the KOPIO 
``catcher''), new veto counters around the beam holes, and new 
high-rate electronics.

Compared to E391a, E14 expects to benefit from tenfold increases in
the $K_L$ rate, signal acceptance, and running time. 
The beamline will be completed and surveyed by 2009. If all 
goes well, the experiment will have an engineering run in 2010, and 
a three-year physics run would begin in 2011.
Three to seven signal events may be observed. Subject to further 
approvals, 
the three-year E14 run could be followed by a series of upgrades to 
the beamline and detector with the goal of pursuing an \order{100}-SM-event 
measurement.
    
An experiment somewhat similar in concept to E391a/E14 has been proposed
at the IHEP 70~GeV synchrotron in Protvino \cite{KLOD+07:prop,OB07:Kaon}. 
The KLOD experiment would use a a neutral beam
extracted at large angle, with an $n/K_L$ ratio of less than 10 
and $10^{8}$ $K_L$ per pulse with a peak momentum of 10~GeV. 
The technical design report for the beamline is complete and R\&D work 
on the detector is ongoing. 
The KLOD experiment would make use of some innovative, cost-contained
technical solutions, particularly in the design of the vetoes. The
main barrel veto uses shashlyk calorimeter modules, with 1~mm
scintillator plates interleaved with 0.3~mm lead foils. 
The downstream calorimeter is a 
lead/scintillating-fiber calorimeter similar in concept to the KLOE EmC, 
with the fibers arranged in $x$-, $u$-, and $v$-views to allow shower 
tracking, providing additional constraints for the 
reconstruction of the $\pi^0$. 
(The KOPIO experiment would have used a preradiator 
in front of the downstream shashlyk calorimeter for shower tracking.)
The in-beam photon veto is also a fiber calorimeter, with dual readout 
(clear and scintillating fibers) to separate electromagnetic and 
hadronic showers. The KLOD experiment has received scientific approval,
but additional funding will be needed for construction and running.

\subsection{More Distant Prospects}
In 2007, because of expected delays in the timeline
for the International Linear Collider (ILC), 
the construction of a high-intensity proton source at Fermilab
was proposed as an interim project \cite{B+07:FermiSGR}. Project X
calls for the construction of an 8 GeV linac with ILC beam 
parameters---essentially a 1:100 scale implementation of the planned
ILC linac technology. The beam from the Project X linac would be accumulated
in the existing Recycler Ring, providing up to 200 kW of fast- or 
slow-extracted 8-GeV proton beam for precision physics ($\mu\to e\gamma$,
kaon physics, and other topics). Beam circulating in the Recycler could 
also be transferred to the Main Injector and accelerated to 120 GeV,
providing 2.3~MW of fast-extracted beam for neutrino physics.

The high intensity of the 8-GeV Project X beam 
(200 kW corresponds to $\sim$\SN{2}{14} ppp for a 1.4~s duty cycle)
would potentially make \order{1000}-SM-event measurements of
$K\to\pi\nu\bar{\nu}$ decays feasible. Experiments with both
charged and neutral kaons have been discussed \cite{A+08:GoldBook}.
The $K^+\to\pi^+\nu\bar{\nu}$ experiment would use stopped kaons and is 
similar in concept to E787/949. The kaon momentum in the secondary
beam is 450 MeV, and a higher field in the central tracker (up to 3~T
in the Project X experiment) makes for a more compact design
with better momentum resolution. The range stack is much more
finely segmented, and a homogeneous liquid-xenon photon veto detector
is used in place of the sampling detectors in E787/949.
The $K_L\to\pi^0\nu\bar{\nu}$ experiment would be conceptually similar to
KOPIO. Like KOPIO, it would make use of a microbunched beam to allow
event-by-event measurement of the momentum of the incident kaon by 
time of flight, as
well as a preradiator to measure the angles of incidence of the photons 
from the $\pi^0$ on the downstream calorimeter. Unlike in KOPIO, which 
used a flat beam in order to increase the beam flux while maintaining
geometrical constraint in at least one dimension, the high Project X 
intensity would allow the use of a tightly collimated and symmetrical beam.
This not only provides good geometrical constraints; it also 
resolves certain technical challenges related to the mechanics of the
KOPIO design.

Unfortunately, in its report on US particle physics priorities for the 
next ten years, the P5 subpanel of the Energy Department's High 
Energy Physics Advisory Panel, while recognizing the importance of 
the above measurements, recommended funding for the kaon physics program
at Project X only in the hypothesis that funding levels for high-energy
physics are doubled over the next ten years \cite{P5+08:10yr}. 
(The subpanel recommended in favor of construction of the Project X 
facility, as well as of other elements of the physics program.)
While the chances of the Project-X kaon physics program being
pursued as proposed are slim, it is perhaps encouraging to see that
world interest in these measurements remains high, and that serious 
discussion of experiments capable of \order{1000}-SM-event measurements 
has started.   

\begin{acknowledgments}
I would like to thank Federico Mescia, Fabio Bossi, Giovanni Mazzitelli, 
David Jaffe, Takeshi Komatsubara, and Bob Tschirhart for helpful discussion
and clarifications, and for reading parts of this manuscript. 
\end{acknowledgments}

\bigskip
\bibliographystyle{elsart-num}
\bibliography{hql08_moulson}

\end{document}